%% file: main.tex
\documentclass[conference]{IEEEtran}
\IEEEoverridecommandlockouts
\usepackage{fancyhdr}
\usepackage{cite}
\usepackage{comment}
\usepackage{amsmath,amssymb,amsfonts}
\usepackage{algorithmic}
\usepackage{graphicx}
\usepackage{textcomp}
\usepackage{xcolor}
\usepackage{array}
\usepackage{mathtools}
\usepackage{booktabs}
\def\BibTeX{{\rm B\kern-.05em{\sc i\kern-.025em b}\kern-.08em
    T\kern-.1667em\lower.7ex\hbox{E}\kern-.125emX}}

\DeclareMathOperator*{\argmin}{arg\,min}

\newcommand{\R}{\mathbb{R}}

\begin{document}

\title{Always be Pre-Training: Representation Learning for Network Intrusion Detection with GNNs}

\author{
    \IEEEauthorblockN{Zhengyao Gu\IEEEauthorrefmark{1}\IEEEauthorrefmark{4}, Diego Troy Lopez\IEEEauthorrefmark{2}\IEEEauthorrefmark{4}, Lilas Alrahis\IEEEauthorrefmark{3}, and Ozgur Sinanoglu\IEEEauthorrefmark{3}}
    
    \IEEEauthorblockA{\IEEEauthorrefmark{1}Center for Data Science, New York University, New York, United States of America\\
        \IEEEauthorrefmark{2}Research Technology Services, New York University, New York, United States of America\\
    \IEEEauthorrefmark{3}New York University Abu Dhabi, Abu Dhabi, United Arab Emirates\\
   {\IEEEauthorrefmark{4}The first two authors contributed equally to this work}\\
                     {Email: \{zg2187, diego.lopez, lma387, os22\}@nyu.edu}}
}

\maketitle
\renewcommand{\headrulewidth}{0.0pt}
\thispagestyle{fancy}
\lhead{}
\rhead{}
\chead{This is the author's version of the work.
The definitive Version of Record will appear in the 2024 International Symposium on Quality Electronic Design (ISQED'24).}
\cfoot{}
\begin{abstract}
Graph neural network-based network intrusion detection systems have recently demonstrated state-of-the-art performance on benchmark datasets. Nevertheless, these methods suffer from a reliance on target encoding for data pre-processing, limiting widespread adoption due to the associated need for annotated labels—a cost-prohibitive requirement. In this work, we propose a solution involving in-context pre-training and the utilization of dense representations for categorical features to jointly overcome the label-dependency limitation. Our approach exhibits remarkable data efficiency, achieving over 98\% of the performance of the supervised state-of-the-art with less than 4\% labeled data on the \texttt{NF-UQ-NIDS-V2} dataset.

\end{abstract}

\begin{IEEEkeywords}Intrusion detection, machine learning, graph neural network, NIDS, few-shot learning, self-supervised learning
\end{IEEEkeywords}

\input{Sections/intro}

\input{Sections/relatedwork}

\input{Sections/background}

\input{Sections/Method}

\input{Sections/Experiments}

\bibliographystyle{IEEEtran}
\bibliography{main.bib}

\end{document}

%% file: Sections/intro.tex
\section{Introduction} \label{sec:intro}
\begin{figure}[!t]
 \includegraphics[width=0.48\textwidth]{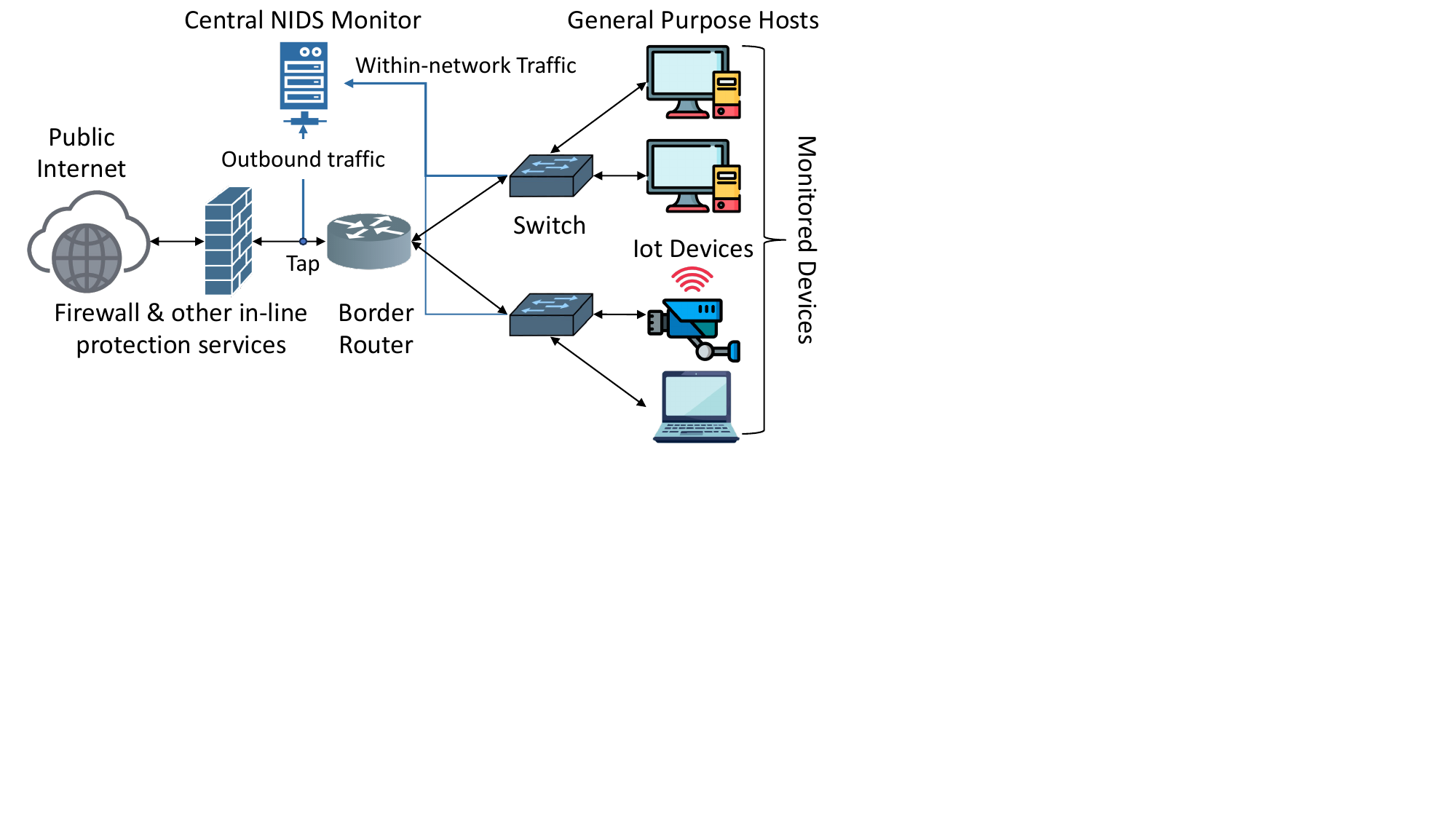}
 \vspace{-4mm} \caption{NIDS deployment, centralizing intrusion detection by monitoring the network of activities~\cite{idssurvey2013}. Monitored devices can be general purpose computer hosts, IoT devices such as IP cameras, wireless devices, or other. Switches move traffic throughout a network and offer mirroring capabilities for NIDS data acquisition. NIDS monitors may be general purpose servers running applications which sniff the mirrored traffic directly on the wire.}

\label{fig:nids}
\end{figure}

A Network Intrusion Detection System (NIDS)\cite{nids} monitors the traffic on a computer network to detect anomalous or malicious activities. 
Undetected intrusions pose threats to the confidentiality, integrity, and availability of computer systems~\cite{idssurvey2013}. Unlike Host Intrusion Detection Systems (HIDS)\cite{hids}, which focus on monitoring system telemetry on individual hosts, NIDS takes a distinct approach. It observes and analyzes the traffic passing through a dedicated point in the network as illustrated in Fig.~\ref{fig:nids}, triggering alerts to downstream threat management when abnormal or suspicious traffic is detected.

With the rise of the Internet-of-Things (IoT), comprised of compute-constrained and low-power devices that are typically unable to run dedicated HIDS, there is a growing demand for centralized intrusion detection systems that run on a separate, dedicated node, such as NIDS~\cite{dastjerdi2016fog}. As can be seen in Fig.~\ref{fig:nids}, in an NIDS system, the monitored devices, including IoT devices, are not engaged in the monitoring process.

A traditional signature-based NIDS compares network forensics to a set of predefined rules and patterns to identify traces that might indicate an attack or intrusion. The rules need to be hand-crafted for each deployment environment and are fundamentally unable to detect novel or zero-day attacks \cite{idssurvey2013}.

More recently, research has focused on intrusion detection as a classification problem in Machine Learning (ML)~\cite{kitsune,bertino2017botnets,dastjerdi2016fog}. ML-based detectors can be trained to recognize baseline patterns in network traffic, enabling them to identify threats beyond known examples. 
Synthetic datasets~\cite{unsw-nb15,ton-iot,cicids,nf-v1,nf-v2} have been utilized to evaluate ML-based solutions for NIDS, as real traffic data is difficult to make public for security concerns~\cite{datasets_not_enough}. 
Additionally, real traffic data contains a minuscule proportion of malicious examples, which can still be cost-prohibitive to label. Such imbalance between benign and malicious activities can also pose challenges to model training~\cite{imbalance}.

Recently, intrusion detectors based on Graph Neural Networks (GNNs)~\cite{egraphsage, anomale, chang2021graphbased, ts-ids} have achieved state-of-the-art performance on synthetic benchmarks.\footnote{In supervised GNN-based NIDS~\cite{egraphsage, concatenated_gnn, node-edge-gnn, early_accurate_nids}, the classifier predicts the labeled class for each training example. In unsupervised methods~\cite{anomale, arganids}, the model learns to encode data distribution without specific label guidance, subsequently using these encodings to establish decision boundaries.} GNN-based NIDS construct graph representations of the unit of activity they monitor. Such graphs can represent low-level activities like packet behavior~\cite{early_accurate_nids} or high-level network flow connections, such as Hypertext Transfer Protocol (HTTP) requests~\cite{ton-iot}. The latter consider topologically related activities (e.g., other network flows sharing a common source or destination node) while detecting intrusion on a single record of network activity (e.g., network flow). This enables the classifiers to better detect coordinated attacks or those involving a chain of actions, such as Distributed Denial of Service (DDoS)~\cite{ddos} and Man-In-The-Middle (MITM)~\cite{mitm} attacks.

\subsection{State-of-the-Art and their Limitations}
While GNN-based NIDS have achieved state-of-the-art performance, they suffer from the following limitations.

\subsubsection{Challenges in Categorical Feature Handling} Many GNN-based NIDS~\cite{egraphsage,anomale,chang2021graphbased} boast state-of-the-art performance on heterogeneous synthetic datasets~\cite{ton-iot,nf-v1,nf-v2}. These dataset are dubbed heterogeneous for their inclusion of features associated with different network protocols such as HTTP, secure sockets layer, and the internet protocol suite (TCP/IP). These features are usually categorical and contain rich semantics for traffic behavior~\cite{ton-iot,big_hetero}. However, current state-of-the-art GNN-based methods either drop or reduce these categorical features to scalar-valued target encodings, which computes the empirical correlation between categories and labels during pre-processing. 
Such practice potentially loses vital information for intrusion detection, which could lead to poor generalization under distribution shift.

\subsubsection{Label Dependency} Obtaining and maintaining an accurately labeled dataset representative of the target domain can be cost-prohibitive, posing a central challenge in NIDS research~\cite{datasets_not_enough}. As an example, recreating a dataset such as \texttt{ToN-IoT}~\cite{ton-iot} would require an organization to identify and label $216,043$ malicious flows within their own environment.

Unsupervised methods offer a solution to the challenge of requiring labeled data for ML model training. However, existing GNN-based unsupervised methods, such as \textit{Anomal-E}~\cite{anomale}, preprocess categorical features using target-encoding. This method computes the empirical correlation between categories and label classes, thus, still relying on labels.

\subsection{Our Contributions}
\textbf{In this study}, we investigate approaches for representing categorical features in GNN-based NIDS to mitigate label-dependency limitations. Initially, we examine the representation of these features using \textit{dense vector embeddings}, as they are more expressive than scalar target encoding. However, we find that directly learning such embeddings through supervised learning leads to poor performance in some settings, evidenced by a performance gap in F1-score of approximately 7\%. Subsequently, we introduce \textit{\textbf{in-context pre-training and fine-tuning}}, a training procedure that initially trains a GNN encoder on unlabeled data from a network before fine-tuning on a small sample of labeled data from the same network, by theorizing that the initial Self-Supervised Learning (SSL) would better initialize the weights before supervised training. We show that this technique not only improves performance compared to target encoded representations but also achieves remarkable data-efficiency.

\textbf{Key-results:} Considering the \texttt{ToN-IoT}~\cite{ton-iot} and \texttt{NF-UQ-NIDS-V2}~\cite{nf-v2} datasets (discussed in Section~\ref{sec:benchmarks}), we observe that the pre-trained model retains over 95\% of its performance on the full dataset when fine-tuned on less than 4\% of the labeled data. In-context pre-training, thus, offers an alternate solution to the label dependency issue through \textit{few-shot learning}. Additionally, pre-trained GNNs can be easily adapted for multi-way classification with a tractable amount of labeled examples, whereas unsupervised methods have only been shown suitable for binary classification.

The paper's structure is as follows: Section~\ref{sec:related_work} summarizes related works in NIDS, and Section~\ref{sec:background} provides an overview of datasets, GNNs, and SSL. Section~\ref{sec:method} introduces in-context pre-training and fine-tuning methods with dense representation, Section~\ref{sec:experiment} details experimental configurations and results, and Section~\ref{sec:conclusion} concludes and discusses future works.

%% file: Sections/relatedwork.tex
\section{Related Works in GNN-based NIDS} \label{sec:related_work}

This section surveys GNN-based NIDS approaches, including supervised and unsupervised methods, and briefly touches on cross-domain NIDS related to label dependence.

\subsection{Supervised GNN-based NIDS} \textit{E-GraphSAGE}~\cite{egraphsage} is the first general-purpose GNN-based NIDS. The method proposed a simple GraphSAGE-like~\cite{graphsage} model that propagates edge feature information across the neighborhood. \textit{E-ResGAT}~\cite{chang2021graphbased} proposed a graph attention network with residual connections operated on line graphs (a graph transformation that convert nodes to edges and vice versa) and showed that the residual connection significantly improved the performance on minority classes. State-of-the-art performance is attained by~\cite{concatenated_gnn}, which proposed a three-layer GNN consisting of two spectral layers as the first and last layer and a spatial aggregation in the middle. \textit{NE-GConv}~\cite{node-edge-gnn} is a GNN-based NIDS that incorporates both node and edge features. In~\cite{spatio-temporal}, the authors considered the temporal evolution of network traffic and proposed a NIDS system based on spatio-temporal GNNs. In~\cite{ts-ids}, classification performance is improved by training a supervised model alongside unsupervised objectives that consider the traffic quantity at each node.

In~\cite{early_accurate_nids}, the authors proposed a \textit{Graph2Vec+} random forest NIDS, performing flow-level classification by constructing graphs for packet burst behavior~\cite{burst}. By focusing on low-level packet behavior rather than extracting statistical features, the system achieves high data efficiency, retaining over 95\% performance with only 10\% of the training data.

\subsection{Unsupervervised GNN-based NIDS} Anomal-E~\cite{anomale} is the first unsupervised GNN-based NIDS. It learns meaningful representations for edges by applying Deep Graph Infomax (DGI)~\cite{dgi}, a contrastive SSL algorithm that maximizes the mutual information between local and global representation of the communication graph. The learned edge embeddings are then fed to traditional unsupervised classification methods, such as isolation forest and Cluster-Based Local Outlier Factor (CBLOF)~\cite{cblof} for intrusion detection. In~\cite{arganids}, the authors proposed \textit{ARGANIDS}, where an Adversarially Regularized Graph Auto-encoder (ARGA)~\cite{arvga} and its variants are first pre-trained on unlabeled data. The learned embeddings from the encoder are then used to train a random forest to classify network flows.

Both approaches~\cite{anomale,arganids} train unsupervised models without subsequent fine-tuning, neglecting the optimization of pre-trained model weights through supervised learning.

\subsection{Cross-Domain NIDS} 
The widespread adoption of supervised ML-based NIDS has been hindered by the overall lack of available data. This is in part due to obstacles, such as organizations being discouraged from sharing their expert-labeled data due to security concerns~\cite{datasets_not_enough}. One strategy to overcome this limitation is to develop ML models that can generalize across various deployment domains. In~\cite{generalizability}, the authors demonstrated that current ML-based NIDS perform poorly across different domains of network traffic. Subsequently,~\cite{diids} proposed training domain-generalizable ML-based NIDS by utilizing only domain-invariant features.

%% file: Sections/background.tex
\section{Background} \label{sec:background}

This section provides essential background information on NIDS dataset benchmarks, GNNs, and SSL. The commonly used notations across the paper are defined in Table~\ref{tab:my_label}.
\subsection{Heterogeneous NIDS Benchmarks} 
\label{sec:benchmarks}
By utilizing various sources of network activities, heterogeneous NIDS datasets allow intrusion detectors to construct a more comprehensive and multi-dimensional understanding of network dynamics, which could allow the NIDS to adapt to new and emerging threats by considering a broad spectrum of indicators and behaviors\cite{ton-iot-hetero}.

\texttt{ToN-IoT}~\cite{ton-iot} was the first to propose a heterogeneous NIDS dataset. \texttt{ToN-IoT}'s heterogeneity is reflected by the inclusion of 40 features across various network protocols. Specifically, the dataset contains connection features from Transmission Control Protocol (TCP) or User Datagram Protocol (UDP) at the transport layer, statistical summaries of TCP/IP activities at internet and transport layers, secure sockets layer states also at the transport layer, and Domain Name System (DNS) and HTTP activities (application layer). \texttt{ToN-IoT} consists of 9 attack types, including backdoor, cross-site scripting (XSS), password cracking, MITM, Denial of Service (DoS), DDoS, scanning, ransomware, and injection attacks. There are $461,043$ total records in the dataset.

\texttt{NF-UQ-NIDS-V2}~\cite{nf-v2} designs 43 standardized NetFlow\cite{netflowv9} features across four datasets: \texttt{UNSW-NB15}~\cite{unsw-nb15}, \texttt{ToN-IoT}~\cite{ton-iot}, \texttt{CSE-CIC-IDS2018}~\cite{cicids}, and \texttt{BoT-IoT}~\cite{bot-iot}. The authors compile \texttt{NF-UQ-NIDS-V2} by extracting the newly designed features from the raw packet capture files from the four aforementioned datasets. Like \texttt{ToN-IoT}, the new feature set contains various network protocols such as TCP, IP, UDP, the Internet Control Message Protocol (ICMP), and secure sockets layer. In addition, the dataset includes other internet and application layer protocols such as DNS and the File Transfer Protocol (FTP). \texttt{NF-UQ-NIDS-V2} contains $75,987,976$ records in total with 20 attack categories.

Earlier datasets, like \texttt{UNSW-NB15}, \texttt{CSE-CIC-IDS2018}, and \texttt{BoT-IoT}, primarily contain statistical features summarizing generic network connections. \texttt{UNSW-NB15} has 40 out of 49 numeric features, while \texttt{CIC-IDS-2018} and \texttt{BoT-IoT} exclusively consist of numeric features. In contrast, \texttt{ToN-IoT} and \texttt{NF-UQ-NIDS-V2} are predominantly composed of categorical features describing protocol activities. \texttt{ToN-IoT} has 31 out of 40 categorical features, and \texttt{NF-UQ-NIDS-V2} has 15 out of 43 categorical features.

\subsection{Graph Neural Networks (GNNs)}

A key feature of GNNs is their capability to learn rich representations of graphs through \textit{message passing} across a graph's topological structure. More specifically, this means combining the representation of each node with an aggregation of neighboring node features to generate a new representation for the node at each layer. This process is usually done multiple times through the different layers of GNN. Formally, this message passing framework is defined by three abstract functions at each layer: \textit{message function} $\phi^{(k)}$, \textit{aggregation operator} $\bigoplus$, and \textit{update funciton} $\gamma^{(k)}$. One can compute the next layer representation by the following formula.
\begin{equation}\label{eqn:message-passing-framework}
 x_u^{(k)} = \gamma^{(k)}\left(x_u^{(k-1)}, \bigoplus_{v \in \mathcal{N}(u)} \phi^{(k-1)}\left(x_u^{(k-1)}, x_v^{(k-1)}, e_{vu}\right) \right)
\end{equation}
Here, $x_u^{(k)} \in \R^{d_x}$ is the hidden representation of node $u$ at the $k$-th layer. Usually, $x_u^{(0)}$ is the node's feature vector associated with node $u$. The message function takes edge features $e_{vu} \in \R^{d_e}$ and the hidden representations of any neighbor $v$ of $u$ as input.
The \textit{aggregation} operator combines all messages from the neighboring nodes $\mathcal{N}(u)$. $\gamma^{(k)}$ is the \textit{update} function that outputs the next layer hidden node representation by taking the aggregated message and the node representation from the previous layer. Different GNNs mainly differ based on the selection of the message, update, and aggregation functions.

\begin{table}[!t]
    \centering
\caption{Commonly used notations}

    \begin{tabular}{cl}
    \toprule
    \textbf{Symbol}     &  \textbf{Definition} \\
    \hline
    $G = (\mathcal{V}, \mathcal{E})$ & Graph $G$ with nodes $\mathcal{V}$ and edges $\mathcal{E}$ \\\hline
    $x^{(k)}_u, u \in \mathcal{V}$     & Vector representation of node $u$ at layer $k$\\\hline
    $x_i, i \in \mathbb{N}$ & A feature in the context of representation learning 
    \\\hline
    $x_{num}, x_{cat}$ & Numeric and categorical input features (encoded) \\\hline
    $h_u$, $h_{uv}$ & Final node/edge representation (embedding)\\\hline
    $h$ & An arbitrary hidden representation\\\hline
    $d_x$, $d_e$ & GNN node/edge hidden dimension \\\hline
    $a$ & Aggregated message\\\hline
    $K$ & Total \# of GNN layers\\\hline
    $e_{uv}, (u, v) \in \mathcal{E}$ & Features of edge $(u, v) \in \mathcal{E}$ \\\hline
    $o_i, i \in \mathbb{N}$ & One-hot encoded vector\\\hline
    $r_{num}$, $r_{cat}$ & Number of numerical and categorical features\\\hline
    $\mathcal{J}(i)$ & The index of $1$ in the one-hot vector $o_i$\\\hline
    $\mathcal{N}(u)$ & The set of all nodes $v$ such that $(v, u) \in \mathcal{E}$ \\\hline
    $\gamma^{(k)}, \phi^{(k)}, \bigoplus$ & Update, message function, aggregation operator\\\hline
    $f_\theta, q_\xi$ & Encoder and decoder in SSL\\\hline
    $\mathcal{D}$ & Data sample\\\hline
    $\mathcal{S}$ & Similarity function in DGI\\\hline
    $\overline{s}$ & Graph summary\\\hline
     $\mathcal{C}(G)$ & Corrupted graph \\\hline
     $\Tilde{h}_{uv}$ & Final edge representation from the corrupted graph\\\hline
     $\omega$, $\varphi$ & Embedding/interaction function \\\hline
    $W, W_1, W_2, W_s$ & Weight matrices \\\bottomrule
    
    \end{tabular}
    \label{tab:my_label}
\end{table}
\subsubsection{E-GraphSAGE} E-GraphSAGE~\cite{egraphsage} encodes edge information in node representations by propagating both node and edge information in its message function. Formally, the message function of E-GraphSAGE is defined as
\begin{equation} \label{eqn:egraphsage-message}
\phi\left(x_u, x_v, e_{vu}\right) = W_1[x_v; e_{vu}],
\end{equation}
where $[x_v; e_{vu}]$ is the concatenation of vector $x_v$ and $e_{vu}$ and $W_1$ is a learnable matrix. Note that the message function output in E-GraphSAGE is invariant to the input central node $x_u$.
The aggregation operation simply averages the messages:
\begin{equation}\label{eqn:egraphsage-agg}
a:= \bigoplus_{v \in \mathcal{N}(u)} \phi_v = \sum_{v \in \mathcal{N}(u)} \phi_v
\end{equation}
Finally, the aggregated message is combined with center node $x_u$ to compute the next layer of representation,
\begin{equation}\label{eqn:egraphsage-update}
\gamma(x_u, a) = \sigma(W_2[x_u; a]),
\end{equation}
where $\sigma$ is a non-linear activation function.
Eq.~(\ref{eqn:egraphsage-message}) and Eq.~(\ref{eqn:egraphsage-update}) together ensure that the updated node representation includes information from the center node representation $x_u$, edge features $e_{uv}$, and the neighboring node representations $x_v$.

In E-GraphSAGE with $K$ layers, the final layer node embeddings $h_u := x_u^{(K)}$ of neighboring node representations are concatenated to form edge embeddings $h_{uv} = [h_u; h_v]$.

\subsection{Graph Self-Supervised Learning (SSL)}
SSL is an ML paradigm that trains models by fitting the inherent structure or characteristics within the data through a self-supervised objective $\mathcal{L}_{SSL}$ without any annotated labels. In the context of network intrusion detection, SSL allows the model to learn meaningful representations from the ubiquitous unlabeled traffic data without explicit labels for different types of intrusions. Graph SSL methods employ and train an encoder $f_\theta$ to transform input data into a latent representation and a decoder $q_\xi$ to convert the latent representation into a training signal on data sample $\mathcal{D}$ with the help of the objective function. The encoder $f_\theta$ can be implemented using GNNs or alternative graph-based learning models, with our emphasis placed on GNN-based encoders. Graph SSL can be formulated as the following optimization problem;

\begin{equation}
\theta^*, \xi^* = \argmin_{\theta, \xi} \mathcal{L}_{SSL}(f_\theta, q_\xi, \mathcal{D}).
\end{equation}

The output embedding from the optimized (i.e., trained) graph encoder $f_{\theta^*}$ can be used as input features for downstream classifiers~\cite{anomale,arganids} or the encoder can be further trained for the same purpose.

Existing GNN-based NIDS research has considered two types of graph SSL techniques: contrastive and generative. Both methods learn to encode the input graph into a lower-dimensional embedding with the encoder $f_\theta$. However, contrastive methods supervise the encoder training with a discriminative decoder $q_\xi$ that maximizes the difference between the embeddings of positive and negative examples, while generative methods do so by using the decoder to reconstruct the input from the embedding. At the time of writing, two SSL-based NIDS have been proposed: Anomal-E~\cite{anomale} and ARGANIDS~\cite{arganids}. The former leverages DGI~\cite{dgi}, a constrastive SSL technique, while the latter employs ARGA (and its variants)\cite{arvga}, a generative SSL technique. Our experiments do not consider ARGANIDS as it is not suitable for reconstructing one-hot input features. Consequently, we only describe Anomal-E in details in this section.

\subsubsection{Anomal-E} DGI learns meaningful representation of graphs by maximizing mutual information between the local and global representations of a graph. In practice, this is achieved by maximizing the similarity $\mathcal{S}$ between the graph summary $\overline{s}_G$ and edge embeddings $h_{uv}$ on the original graph $G$ and minimizing the similarity between the same graph summary and the edge embeddings of a corrupted graph $\mathcal{C}(G)$. Specifically, the encoder is trained using gradient descent with the following binary cross-entropy loss objective function.

\begin{equation}\label{eqn:anomale-obj}
\begin{aligned}
\mathcal{L}_{\text{Anomal-E}} = \frac{1}{|\mathcal{E}|} \sum_{(u, v) \in \mathcal{E}} & \log\left( \mathcal{S}(h_{uv}, \overline{s}_G)\right)\\
& + \log\left(1 - \mathcal{S}(\Tilde{h}_{uv}, \overline{s}_{G})\right),
\end{aligned}
\end{equation}

The edge embeddings $h_{uv}$ and $\Tilde{h}_{uv}$ are outputs from an E-GraphSAGE encoder $f_\theta$ on the original graph $G$ and corrupted graph $\mathcal{C}(G)$, respectively. The graph summary $\overline{s}_G$ is defined as the sum of edge embeddings of the original graph $G$:

\begin{equation}
\overline{s}_G = \sum_{(u, v) \in G} h_{uv}
\end{equation}

The corruption function $\mathcal{C}$ in Anomal-E randomly shuffles the edge features among edges of the input graph. The corrupted graph serves as an negative example that does not belong to the input distribution. By optimizing the objective in Eq.~(\ref{eqn:anomale-obj}), Anomal-E learns to separate the embeddings of original graphs and corrupted graphs in the Euclidean space. Such method uses the \textit{contrast} between self-generated examples as training signal, thus dubbed contrastive methods.

Finally, the similarity function between the graph summary $\overline{s}_G$ and the graph embedding $h_{uv}$ in Anomal-E is defined as

\begin{equation}
\mathcal{S}(h_{uv}, \overline{s}_G) = \sigma(h_{uv} W_s \overline{s}^\top_G)
\end{equation}
where $W_s$ is a trainable matrix.

After training the encoder (E-GraphSAGE for Anomal-E), test embeddings are generated from the tuned model for use in unsupervised downstream anomaly detection algorithms, without further fine-tuning.

%% file: Sections/Method.tex
\begin{figure*}[t]
 \centering
 \includegraphics[width=0.78\textwidth]{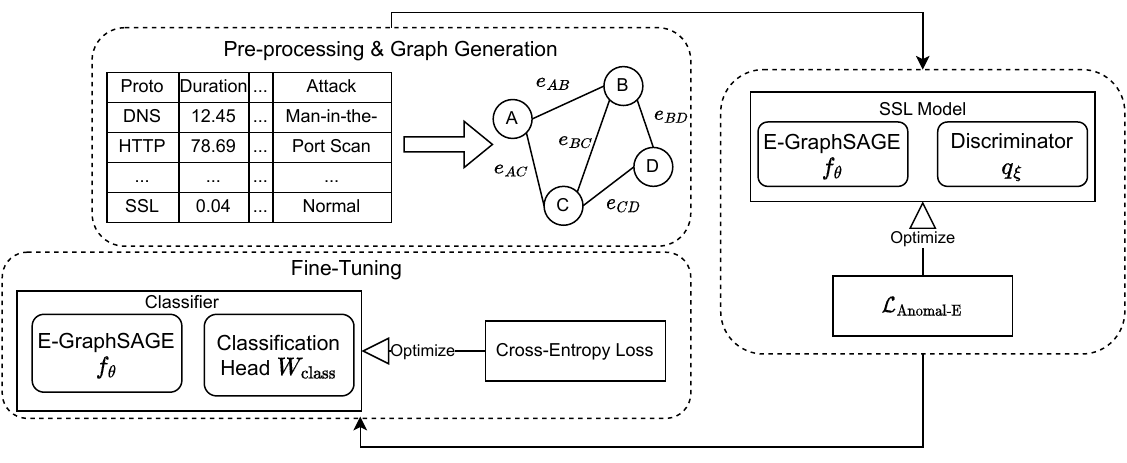}
 \vspace{-15pt}
 \caption{The proposed training pipeline with in-context pre-training, demonstrated using Anomal-E~\cite{anomale} as the SSL technique. Here $A, B, C, D$ represent four arbitrary nodes (flow endpoints). The edges between them represent the flows. During the pre-training phase, we train an encoder $f_\theta$ and a decoder $q_\xi$. Subsequently, $f_\theta$ is connected to a classification head and trained using labeled data.}
 \label{fig:in-context-pre-training}
\end{figure*}

\section{Proposed Method}\label{sec:method}

Here, we first define dense vector representation within the context of intrusion detection and provide examples of implementing such representations for both numeric and categorical features. Subsequently, we discuss the necessity of in-context pre-training for dense representation models.

\subsection{Dense Vector Representation}
Recent heterogeneous NIDS datasets include high-level protocol features, which contain rich semantics that describe network dynamics. Consider the following DNS features in a \texttt{ToN-IoT} example~\cite{ton-iot}, describing a DNS request for the IPv4 address of Google home page that is not rejected and successfully returned: \texttt{dns\_query("google.com")}, \texttt{dns\_qtype("A")}, \texttt{dns\_rcode("NOERROR")}, \texttt{dns\_rejected(FALSE)}.

 The features contain individual semantic meanings, while construct a more contextual meaning together. To capture both layers of meaning, we propose learning an \textit{embedding} function $\omega$, which maps each feature to a high-dimensional vector representation, and an \textit{interaction} function $\varphi$, which models the dynamics between features.

An embedding function of feature $x_i$ is a function $\omega_i: \mathcal{X}_i \xhookrightarrow{} \R^{d_i}$ from the feature space $\mathcal{X}_i$ to the Euclidean space. $d_i$ is the size of individual feature embeddings. Define $\omega(x_1, x_2, \dots, x_r) = (\omega_1(x_1), \omega_2(x_2), \dots, \omega_k(x_r))$. An interaction function $\varphi: \R^{r \times d_i} \to \R^{d_e}$ takes all embedded features $\{\omega_i(x_i)\}_{i \geq 1}$ as input and combines them to output a single vector of size $d_e$. In ML literature, the composite function $\varphi\circ \omega$ is often parameterized jointly and referred to as a feature map. We parameterize them separately as they serve different purposes. Next, we introduce two common practices in representation learning for embedding numeric and categorical feature respectively.

\subsubsection{Linear Projection on Numeric Features.} Let $x \in \R^{r_{num}}$ be a vector with $r$ numeric features and trainable weights $W \in \R^{d_e \times {r_{num}}}$. The linear projection $\mathbf{x} = Wx$ defines an embedding function and an interaction function. Let the $i$-th entry of the $j$-th feature $[\omega_j(x_j)]_i = W_{ij} x_j$, we have

\begin{equation}
x_{num} = Wx = \sum_{j} W_{ij} x_j = \sum_{j=1}^{r_{num}}\omega_j(x) = \varphi(\omega(x))
\end{equation}

We see that linear projection embeds each feature $x_j$ as the $j$-th column $W_{:,j}$, scaled by $x_j$, and combines each embedded feature through summation. Next, we show that projecting one-hot encoding with a linear transformation yields similar embedding and interaction for categorical features.

\subsubsection{One-Hot Encoding for Categorical Features} Suppose we aim to encode a vector of categorical features $x=(x_1, x_2, \dots, x_{r_{cat}}) \in \mathcal{X}_1 \times \mathcal{X}_2 \times \dots \times \mathcal{X}_{r_{cat}}$, where $|\mathcal{X}_i| = c_i$ is the number of possible categories for feature $i$. A common method for encoding these features involves converting each feature into a one-hot vector and then transforming the vector using a linear projection:

$x_{cat} = Wo$, where $o = [o_1^\top; o_2^\top; \dots; o_{r_{cat}}^\top]^\top$ is a concatenation of one-hot vectors $o_i$ of length $c_i$ and $W \in \R^{d_e \times c}$. $c = \sum_{i = 1}^{r_{cat}} c_i$. We claim that the transformation defines an embedding function and an interaction function. 

Write the weight matrix $W$ as the concatenation of a column of smaller matrices $[W_1; W_2; \dots; W_{r_{cat}}]$ where $W_i \in \R^{d_e \times c_i}$. By linear algebra, we have; $\mathbf{x}_{cat} = \sum_{i=1}^{r_{cat}} W_i o_i$.

Let $\mathcal{J}(i)$ indicate the index of $1$ in the one-hot vector $o_i$.
\begin{equation}
x_{cat} = \sum_{i=1}^{r_{cat}} W_i o_i = \sum_{i=1}^{r_{cat}} (W_i)_{:, \mathcal{J}(i)},
\end{equation}
where $(W_i)_{:, \mathcal{J}(i)}$ is the $\mathcal{J}(i)$-th column of matrix $W_i$. Thus, each feature $i$ is embedded as a vector, $\omega_i(x_i) = (W_i)_{:, \mathcal{J}(i)}$, and combined through addition.

\subsubsection{Mixed Features} From the two examples above, we observe that both numeric and categorical features can be embedded as vectors through a linear projection. When the input features comprise both numeric and categorical types, we can define an interaction function that integrates both. In our approach, we generate the final vector representation of the features by concatenating the numeric and categorical embeddings obtained previously, and then pass them through a linear projection:$h = W [x_{num}; x_{cat}]$.

\subsection{In-Context Pre-training}
Self-supervised pre-training has been shown effective in few-shot learning\cite{few-shot-learning} and improving model generalization\cite{ssl_generalization}. SSL can leverage the abundant unlabeled network traffic data for pre-training. This allows the model to learn from the inherent structure and patterns in network dynamics without the need for explicit labels. During pre-training, SSL objectives encourage the model to understand relationships and context within the raw network traffic, as opposed to the its correlation with labels. This leads to the learning of generic representations that capture essential features and information, making the model more capable of generalizing to unseen instances. We show later in experiments that this ability to learn generalizable weights is essential to applying dense representation to GNN-based NIDS.

Conventional pre-training typically involves obtaining data from a broader domain than that of the fine-tuning task. However, defining a broader domain for intrusion detection on a specific network is challenging. Instead, we propose \textbf{in-context pre-training}, where the pre-training data is sourced from the same network as the labeled data. 

\subsection{Fine-tuning}
Fig.~\ref{fig:in-context-pre-training} illustrates the proposed process of training an intrusion detector with in-context pre-training. The process consists of two training stages: (i) SSL pre-training and (ii) fine-tuning the SSL encoder. During pre-training, both the weights of the encoder and the decoder are updated by an SSL objective. The weights of the encoder are later further trained during fine-tuning using supervised objectives. We leave the detailed account of pre-processin to Section \ref{sec:experiment}. This approach results in an ML-based NIDS methodology that is less reliant on labeled data. To train an intrusion detector for a specific network from scratch, one first compiles a representative dataset consisting of raw, unlabeled traffic from the network. After labeling a manageable subset of the traffic (see Section~\ref{sec:experiment} for details on the effort involved), one obtains a ready-to-use intrusion detector through fine-tuning.

%% file: Sections/Experiments.tex
\section{Experiments}\label{sec:experiment}
Next, we outline our experimental configurations and discuss the experimental results. We first introduce the datasets on which we evaluate our methods. Then we detail our data pre-processing method. Finally, we motivate our experiments by enumerating three research questions, which are answered in conjunction with an analysis of the experimental results.

\subsection{Datasets}
We choose the \texttt{ToN-IoT}~\cite{ton-iot} and \texttt{NF-UQ-NIDS-V2}~\cite{nf-v2} datasets due to their rich categorical features. \texttt{ToN-IoT} comprises 65\% normal flows and 35\% malicious traffic, while \texttt{NF-UQ-NIDS-V2} contains 33.12\% and 66.88\% normal and malicious flows, respectively. \texttt{NF-UQ-NIDS-V2}, compiled from four simulated attack scenarios, presents a more challenging scenario than \texttt{ToN-IoT} in our experiments.\footnote{We refer the readers to \cite{ton-iot} and \cite{nf-v2} for the description of the features in \texttt{ToN-IoT} and \texttt{NF-UQ-NIDS-V2}, respectively.} Due to memory limitations, we sub-sample both datasets to around 130,000 training and 50,000 testing flows each. We under-sample the majority classes to balance the label distribution.

\subsection{Hardware Configuration}
We run all experiments on an RTX 3070 GPU with 8GB of memory. We use 16GB RAM and an Intel I5-12600K CPU. The model was also tested solely on the CPU during inference, making it readily deployable on typical NIDS hardware.

\subsection{Pre-processing \& Graph Construction.}
We utilize the standard features provided by~\cite{ton-iot} and~\cite{nf-v2}. Source port numbers are omitted from the feature set. Consistent with~\cite{egraphsage}, we employ uniformly randomly generated IPs in the IPv4 range to represent the source of the flows. This encompasses any address from \texttt{0.0.0.0} to \texttt{255.255.255.255}. We identify the flow destinations using a combination of IP addresses and port numbers.

We pre-process the dataset leveraging various Python packages. Numeric features are standardized with \texttt{StandardScaler} in \texttt{scikit-learn} directly unless the standard deviation of a feature is twice larger than the mean, in which case we take the binary logarithm of the feature first. All unbounded categorical features are dropped, except for ``DNS query'' in \texttt{ToN-IoT}, which we convert to a binary feature that indicates whether the query belongs to Majestic Million top one million most visited domains\footnote{https://majestic.com/reports/majestic-million} at the time of research. Subsequently, the remaining categorical features, along with the added binary feature, are converted to one-hot vectors as described in Section \ref{sec:method}.

We represent hosts in the network as nodes and flows between them as edges in the graph. In line with~\cite{anomale}, the graphs are constructed as undirected multi-graphs. The numeric features and categorical features are concatenated and stored as edge features. The node features of the graph are initialized as vectors of 1's with a length of 64.

\begin{figure}[!t]
 \centering
 \includegraphics[width=0.47\textwidth]{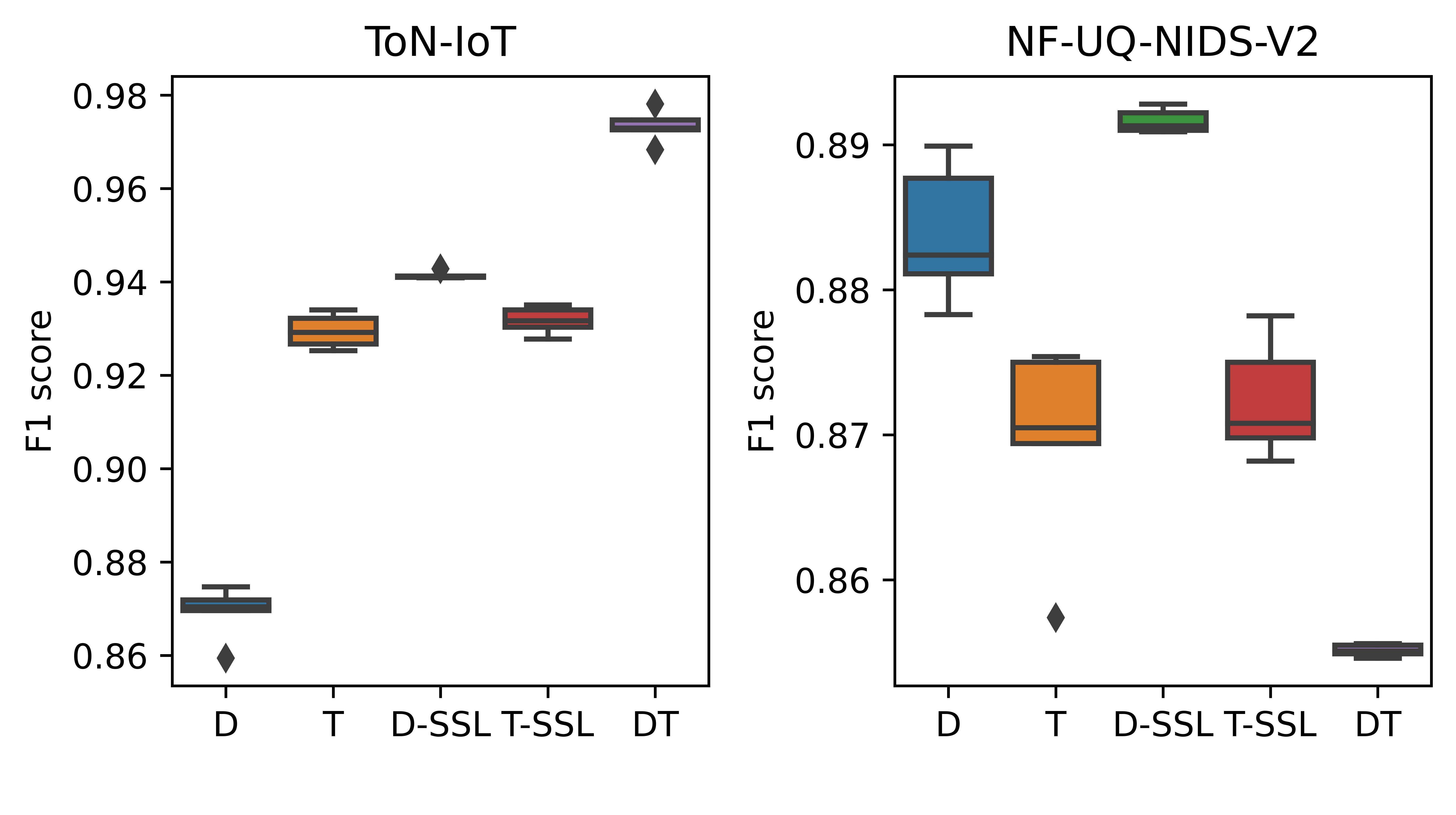}
 \vspace{-12pt}
 \caption{Full-data setting: F1-score of dense representation models (D) and target encoding models (T) trained on all labeled data. For each model type, we experiment with a SSL pre-trained variant (-SSL) and one without. Decision Tree (DT) is included here as a baseline.}
 \label{fig:full-data}
\end{figure}
\subsection{Evaluation Metric}
We evaluate intrusion detectors with F1-score, defined as: $\mathrm{F1} = 2 \times \frac{\mathrm{Precision} \times \mathrm{Recall}}{\mathrm{Precision} + \mathrm{Recall}}$, where Precision represents the rate of correctly identifying true instances among alerts, while Recall indicates the percentage of true instances correctly alerted among all true instances. F1-score, as the harmonic mean of Precision and Recall, provides a summary of both abilities. We report the micro-F1-score, which aggregates F1-scores across classes using a weighted average based on the number of instances in each class, summarizing the overall performance on a specific dataset.

\subsection{Research Questions and Experimental Settings}
Our experiments aim to answer the following questions;

\noindent\textbf{Q1.}~How does dense representation models perform compared to target encoding models?

\noindent\textbf{Q2.}~Does in-context pre-training enhance GNN performance?

\noindent\textbf{Q3.}~Does in-context pre-training result in a high-performing NIDS solution even when labeled training data is scarce?

The research questions are answered in two experimental settings, the \textit{full-data setting} and the \textit{few-shot setting}. In the full-data setting, models are trained on all the available labeled data. As we find the single F1-scores in this setting can be close across different models, we report the F1-scores of 5 trials for each model variant. We present the trial results in a box plot. In the few-shot setting, the models are provided with at most 5,000 labeled examples, and further benchmarked with a decreasing amount of labeled data. We sub-sample the original training data with an under-sampler, which resulted in a more balanced label distribution. This shows that manual crafting of small datasets, typically results in balanced classes.

\subsection{GNN Architecture and Hyperparameters}
Across our experiments, we use E-GraphSAGE as the backbone of the GNNs with two variation factors: 1) whether the GNN embeds the input features with dense representation or target encoding, and 2) whether E-GraphSAGE is pre-trained with Anomal-E as an encoder. We use acronyms D and T to represent the GNNs without pre-training that utilize either dense representation or target encoding embeddings respectively, and D-SSL and T-SSL to refer to their SSL pre-trained counterparts. In addition, we include Decision Tree (DT) as a baseline. In the the full-data setting, we benchmark the five aforementioned models on both datasets. In the few-shot setting, we only include the best-performing unpre-trained GNN from the full-data setting as the representative.

We train all the GNNs using the Adam optimizer with a fixed learning rate of 0.001. Each GNN architecture comprises two layers with a hidden dimension of 128, resulting in an output edge representation size of 256. We train the \texttt{sklearn} implementation of DT using the \texttt{gini} criterion and set \texttt{max\_splitting\_size} to 5. This set of hyperparameters are picked to maximize validation performance.

\subsection{Results and Discussion}
\subsubsection{Full-Data Results Fig.~\ref{fig:full-data}} D-SSL is the best-performing among GNNs. SSL improves the dense representation model on both datasets. It increases the mean F1-score by 8.32\% and 0.87\% and decreases the the variance by 82.45\% and 29.73\% on \texttt{ToN-IoT} and \texttt{NF-UQ-NIDS-V2}, respectively. Meanwhile, there is no consistent improvement in the mean or variance of the F1-score for the target encoding models (T) when combined with SSL across the two datasets. 

Dense representation models performed comparatively better on \texttt{NF-UQ-NIDS-V2} than the other models, over-performing DT by 4.27\% on \texttt{NF-UQ-NIDS-V2} in mean F1-score compared to the 3.28\% underperformance on \texttt{ToN-IoT}. This phenomenon can be explained by the fact that DT and target encoding models, unparameterized and less parameterized than dense representation models, have an advantage on simpler datasets like \texttt{ToN-IoT}, but struggle more on more complex datasets like \texttt{NF-UQ-NIDS-V2}.

\subsubsection{Few-Shot Results}

\begin{figure}
\centering
\includegraphics[width=0.4\textwidth]{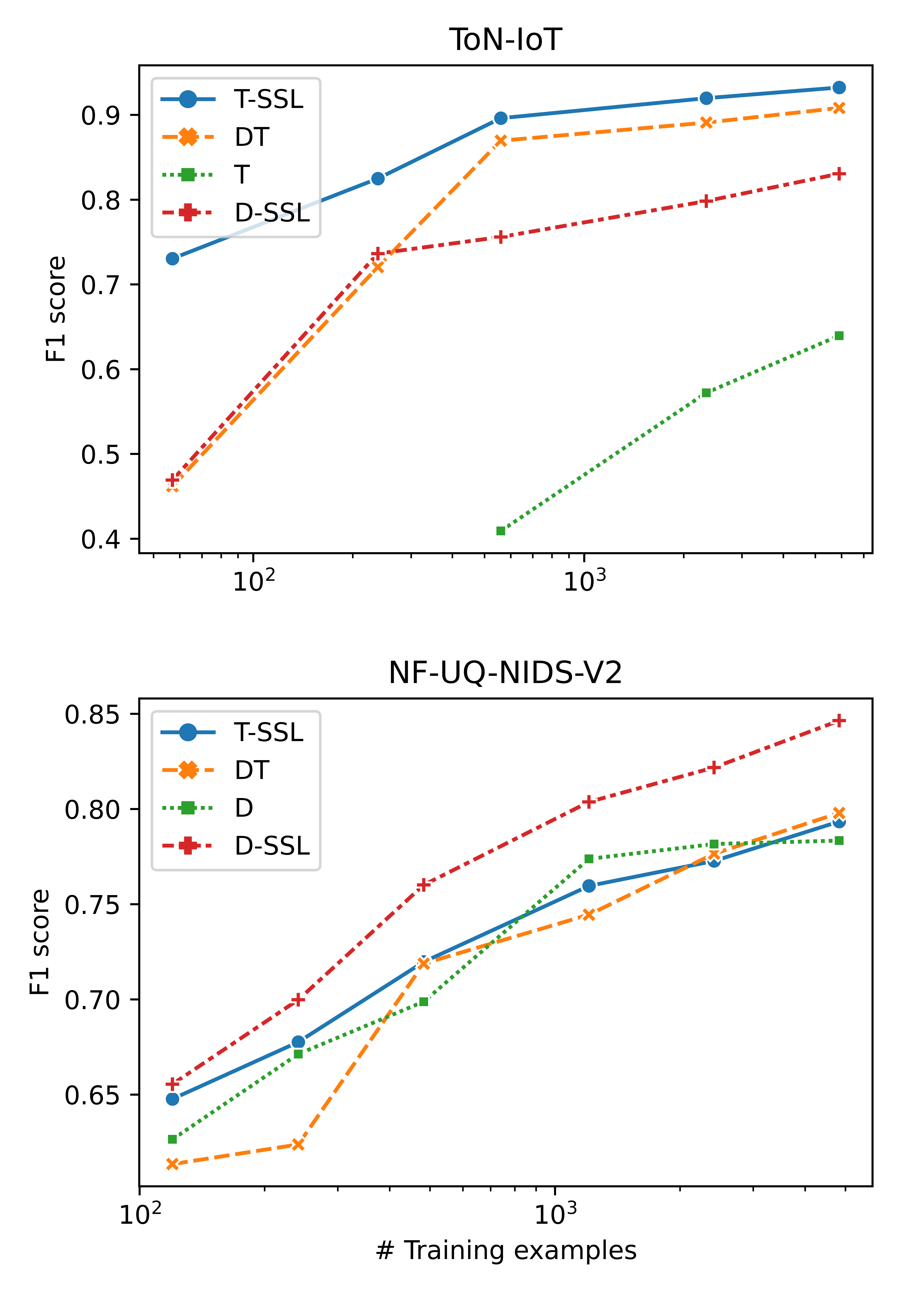}
\vspace{-12pt}
\caption{Few-shot setting: Comparing pre-trained (target encoding T-SSL, dense representation D-SSL) and directly learned models (E-GraphSAGE, Decision Tree DT) on \texttt{ToN-IoT} (top) and \texttt{NF-UQ-NIDS-V2} (bottom) with limited data. We present the best-performing E-GraphSAGE without pre-training in the plots (either target encoding T or dense representation D).}
\label{fig:limited data}
\end{figure}

The results in Fig.~\ref{fig:limited data} depict the performance in the few-shot learning setting. Across both datasets, pre-trained models demonstrate the greatest resilience to data scarcity. Remarkably, the top-performing models, T-SSL and D-SSL, achieved F1-scores of 93.22\% and 84.64\%, respectively. These scores correspond to 99.8\% and 95.11\% of the highest F1-scores attained in the full-data setting, using only 3.7\% of the original training data.

DT ranks second on both datasets with over 500 labeled examples, while the second best pre-trained model ranks second with less than 500 labeled examples. This suggests that pre-trained models excel in learning robust representations despite data scarcity. Moreover, DT outperforming some pre-trained GNNs raises questions about the adequacy of current synthetic datasets for challenging ML-based NIDS.

\subsubsection{Summary of Findings}
Combining the results from the two experiments, we can answer our research questions;

\noindent\textbf{A1.}~While the dense representation GNNs outperform target encoding GNNs on \texttt{NF-UQ-NIDS-V2}, target encoding proves superior on the simpler dataset without pre-training. However, with in-context pre-training, dense representation GNNs consistently outperform target encoding models.

\noindent\textbf{A2.}~While in-context pre-training is typically avoided in other deep learning tasks due to concerns about overfitting, it consistently enhances the performance of dense representation GNNs on our considered NIDS datasets.

\noindent\textbf{A3.}~In our experiments, SSL pre-trained models exhibit significantly greater data efficiency compared to other methods, demonstrating their superiority by a large margin.

\section{Conclusion and Future Work} \label{sec:conclusion}
We have shown that, when pre-trained using SSL, dense representation is the superior method to embed features for GNN-based NIDS. In addition, we have demonstrated that in-context pre-training yields data-efficient models that attain comparable performance with just a fraction of training data compared to previous state-of-the-art models. By proposing a few-shot learning framework, we add to the body of work that addresses the label-dependence of ML-based NIDS.

This work only considers the most basic GNN configurations and graph SSL techniques. We leave it for future work to explore more advanced combination of GNN architectures and graph SSL techniques. Moving forward, we plan to evaluate our framework using real network traffic to gather informative insights into model performance on more realistic and complex distributions. It is also vital for future work to incorporate the directionality of network flows in future GNN-based NIDS.